\documentclass[twocolumn,showpacs,preprintnumbers,amsmath,amssymb,floatfix]{revtex4}

\usepackage{graphicx}
\usepackage{dcolumn}
\usepackage{bm}

\newcommand{\beq}{\begin{equation}}
\newcommand{\eeq}{\end{equation}}
\newcommand{\bey}{\begin{eqnarray}}
\newcommand{\eey}{\end{eqnarray}}

\begin{document}

\title{Transitioning scenario of Bianchi-I universe within 
$f\left(R,T\right)$ formalism}
\author{Anil Kumar Yadav}
\email{abanilyadav@yahoo.co.in} \affiliation{Department of
Physics, United College of Engineering and Research, Greater Noida - 201310, India}

\begin{abstract}
In this paper, we report the existence of transitioning scenario of Bianchi
type I universe in the context of $f\left(R,T\right)$ gravity with special case $f\left(R,T\right) = f_{1}(R) + f_{2}(R)f_{3}(T)$ and it's functional forms $f_{1}(R) = f_{2}(R) = R$ and $f_{3}(T) = \alpha T$ with $\alpha$ being constant. The exact solution of the Einstein's field equations are derived by using the generalized hybrid expansion law that yields the model of transitioning universe from early deceleration phase to current acceleration phase. Under this specification, 
we obtain the singular as well as non singular solution of Bianchi type I model depending upon the particular choice of the value of problem parameters. We also notice that the validation of weak energy condition and dominant energy condition and violation of strong energy condition occurs for these values of problem parameters. The deceleration parameter is found to be negative at present (z = 0) in the derived model which is supported by recent observations.\\    

\end{abstract}



\maketitle

\section{Introduction}
\label{intro}
In order to explain the cosmic acceleration, $f(R,T)$ theory of gravity \cite{Harko/2011} are an optimistic alternative of general relativity with cosmological constant $(\Lambda)$ because GR with $(\Lambda)$ faces  some problems on theoretical ground like cosmic coincidence and fine tuned. The Refs. \cite{Riess/1998,Perlmutter/1999} have evidence that the model of accelerating universe without cosmological constant $(\Lambda)$ is possible only when about $95\%$ energy/matter contents of the present universe is in the form of dark energy/matter \cite{Kumar/2017}. However the nature of dark energy/matter is still dubious \cite{Dil/2017,Behrouz/2017,Germani/2017,Jennen/2016}. In $f(R,T)$ gravity, the matter Lagrangian is coupled with Ricci scalar $(R)$ and trace of energy momentum tensor $(T)$ \cite{Yadav/2014,Moraes/2017,Yadav/2018}.
The $T$ - dependence $f(R,T)$ gravity leads to the possibilities of consideration of quantum effects that yields the
probabilities of production of particles \cite{Parker/2014}. Such possibilities may give a clue that there is a connection
between quantum theory of gravity and extended theory of gravity with matter-geometry coupling.\\

Despite of recent elaboration, the $f(R,T)$ theory of gravitation has already been applied in Astrophysics \cite{Zubair/2016,Yousaf/2017,Moraes/2017r,MoraesSahoo/2017,Das/2016,Sharif/2014} as well as in Cosmology \cite{Yadav/2014,Yadav/2015,Yadav/2018,Moraes/2017,Myrazakulov/2012,Singh/2015}. Shabani and Farhoudi \cite{Shabani/2014} have deliberated the cosmological and solar system consequences of $f(R,T)$ gravity model. The Refs. \cite{Shabani/2014} deals with parametrized post-Newtinian parameter for $f(R,T)$ theory of gravitation and shows that this model may accept admissible values of parametrized post-Newtinian parameter especially 1 in case of $f(R,T) = f_{1}(R) + f_{2}(T)$. In the recent past, Kiani and Nozari \cite{Kiani/2014} have studied $f(R,T)$ model based on the scalar perturbation in the space-time. In \cite{Zubair/2016,Yousaf/2017,Moraes/2017,MoraesSahoo/2017} the authors have investigated wormhole solution in $f(R,T)$ gravity. Moraes and Sahoo \cite{Moraes/2017} have deliberated non-minimal matter-geometry coupling, governed by hybrid expansion law for isotropic and homogeneous universe. Our main goal in the present
paper is to develop the anisotropic non-minimal matter geometry coupling in $f(R,T)$ theory of gravity by taking into account the generalized hybrid expansion law \cite{Yadav/2013,Ozgur/2014} that generates transitioning model of universe and describes both the early decelerated phase and present accelerated phase of universe expansion as well as transition between these two regimes. Shen and Zhao \cite{Shen/2014} have studied quintom model of universe in $f(R,T)$ gravity which gives the periodic varyiation of deceleration parameter. In the Ref. \cite{Aygun/2016}, the authors have elaborated the the idea of accelerating universe without attribution of dark energy/dark mater in $f(R,t)$ theory of gravitation and show that the acceleration is the geometrical property of the present universe. 
Recently, Moraes et al \cite{Moraes/2017a} have searched a cosmological scenario from the Starobinsky model with in the framework of $f(R,T) = R + \alpha R^{2} + 2\gamma T$ formalism with $\alpha$ and $\gamma$ being constant. In the present paper we will focus our attention on $f(R,T) = f_{1}(R) + f_{2}(R)f_{3}(T)$ gravity which has shown to provide an alternative to the cosmological issues like cosmic coincidence and fine tune in general theory of relativity. Also, the Refs. \cite{Moraes/2017,Zaregonbadi/2016} give an evidence that $f(R,T)$ gravity does not need to invoke the dark energy/dark matter for acceleration.\\ 

A Bianchi type I universe is the simplest and straightforward generalization of FRW universe because it describes homogeneous and anisotropic universe with different scale factors
along each spatial directions. Despite of recent elaboration, in the literature, several authors have
studied Bianchi type I model in different physical contexts \cite{Akarsu/2010,Kumar/2011,Yadav/2012,Saha/2004,Yadav/2016}. In the present paper, we confine ourselves to study the non-minimal matter geometry coupling
governed by generalized hybrid expansion law within 
$f\left(R,T\right)$ formalism . The paper is organized as follow: in section 2, we formulate basic mathematical formalism of $f\left(R,T\right) = f_{1}(R) + f_{2}(R)f_{3}(T)$ gravity with specific
choice i.e. $f_{1}(R) = f_{2}(R) = R$ and $f_3(T) = \alpha T$. In section 3, we have computed the field equations
for $f\left(R,T\right) = R + \alpha RT$ theory of gravitation in Bianchi type I space-time. The
physical consequences of the derived model and energy conditions have been discussed in section 4.
In section 5, we have concluded our results.  

\section{The $f\left(R,T\right) = f_{1}(R) + f_{2}(R)f_{3}(T)$ Gravity}
\label{sec:1}
The $f\left(R,T\right)$ theory of gravity is the
modification of general relativity (GR). The action for $f(R,T)$ theory
is given by \cite{Harko/2011,Yadav/2014}:
\begin{equation}
\label{FRT1}
S = \int \sqrt{-g} \left(\frac{f(R,T)}{16 \pi G} + L_{m}\right)dx^{4},
\end{equation}
where $f\left(R,T\right)$ is an arbitrary function of the Ricci
scalar $R$ and the trace $T$ of energy momentum tensor $T_{ij}$
while $L_m$ is the usual matter Lagrangian.\\ 
The energy momentum tensor $T_{ij}$ is read as
\begin{equation}
 \label{FRT2}
T_{ij}\,=\,-\frac{2}{\sqrt{-g}}\,\frac{\delta\left(\sqrt{-g}\,L_m\right)}{\delta
g^{ij}}
\end{equation}
The gravitational field of $f(R,T)$ gravity is given by
\begin{equation}
\label{basic1}
\begin{array}{ll}
\left[f_{1}^{\prime}(R)+f_{2}^{\prime}(R)f_{3}^{\prime}(T)\right] R_{ij}-\frac{1}{2}f_{1}^{\prime}(R)g_{ij}+
\left(g_{ij}\bigtriangledown^{i} \bigtriangledown_{i}  - \bigtriangledown_{i} \bigtriangledown_{j}\right)\\
\\
\,\,\,\,\,\,\,
\times \left[f_{1}^{\prime}(R)+f_{2}^{\prime}(R)f_{3}^{\prime}(T)\right] = \left[8\pi + {f_{2}}^{\prime}(R){f_{3}}^{\prime}(T)\right]T_{ij}+\\
\\
\,\,\,\,\,\,\,\,\,\,
f_{2}(R)\left[f_{3}^{\prime}(T)p+\frac{1}{2}f_{3}(T)\right]g_{ij}
 \end{array}
\end{equation}
Here, $f(R,T) = f_{1}(R)+f_{2}(R)f_{3}(T)$ and primes denote derivatives with respect to the arrangement. We assume $f_{1}(R) = f_{2}(R) = R$ and $f_{3}(T) = \alpha T$, with
$\alpha$ as a constant \cite{Moraes/2017}.\\
Thus the equation (\ref{basic1}) yields
\begin{equation}
 \label{basic2}
G_{ij} = 8\pi T^{(eff)}_{ij} = 8\pi(T_{ij}+T^{(DE)}_{ij})
\end{equation}
where, $T^{(eff)}_{ij}$, $T_{ij}$ and $T^{(DE)}_{ij}$ represent the effective energy momentum tensor, matter energy momentum tensor and dark energy term respectively. The dark energy term is read as
\begin{equation}
 \label{basic3}
T^{(DE)}_{ij}=\frac{\alpha R}{8\pi}\left(T_{ij}+\frac{3\rho-7p}{2}g_{ij}\right).
\end{equation}
It is worth to note that the term $T^{(DE)}_{ij}$ arises due the matter-energy coupling in present theory \cite{Moraes/2017}.\\
By applying the Bianchi identities in equation (\ref{basic2}) yields
\begin{equation}
 \label{basic4}
\bigtriangledown^{i}T_{ij} = -\frac{\alpha R}{8\pi}\left[\bigtriangledown^{i}(T_{ij}+pg_{ij})+
\frac{1}{2}g_{ij}\bigtriangledown^{i}(\rho-3p)\right]
\end{equation}
\section{The metric and field equations}
The line element of Bianchi type I space-time is read as
\begin{equation}
 \label{spacetime}
ds^{2}=dt^{2}-A^2\,dx^{2}-B^2\,dy^{2}-C^2\,dz^2,
\end{equation}
where $A=A(t)$, $B=B(t)$ and $C=C(t)$ are functions of $t$ only.\\
\begin{figure*}[thbp]
\begin{tabular}{rl}
\includegraphics[width=7.5cm]{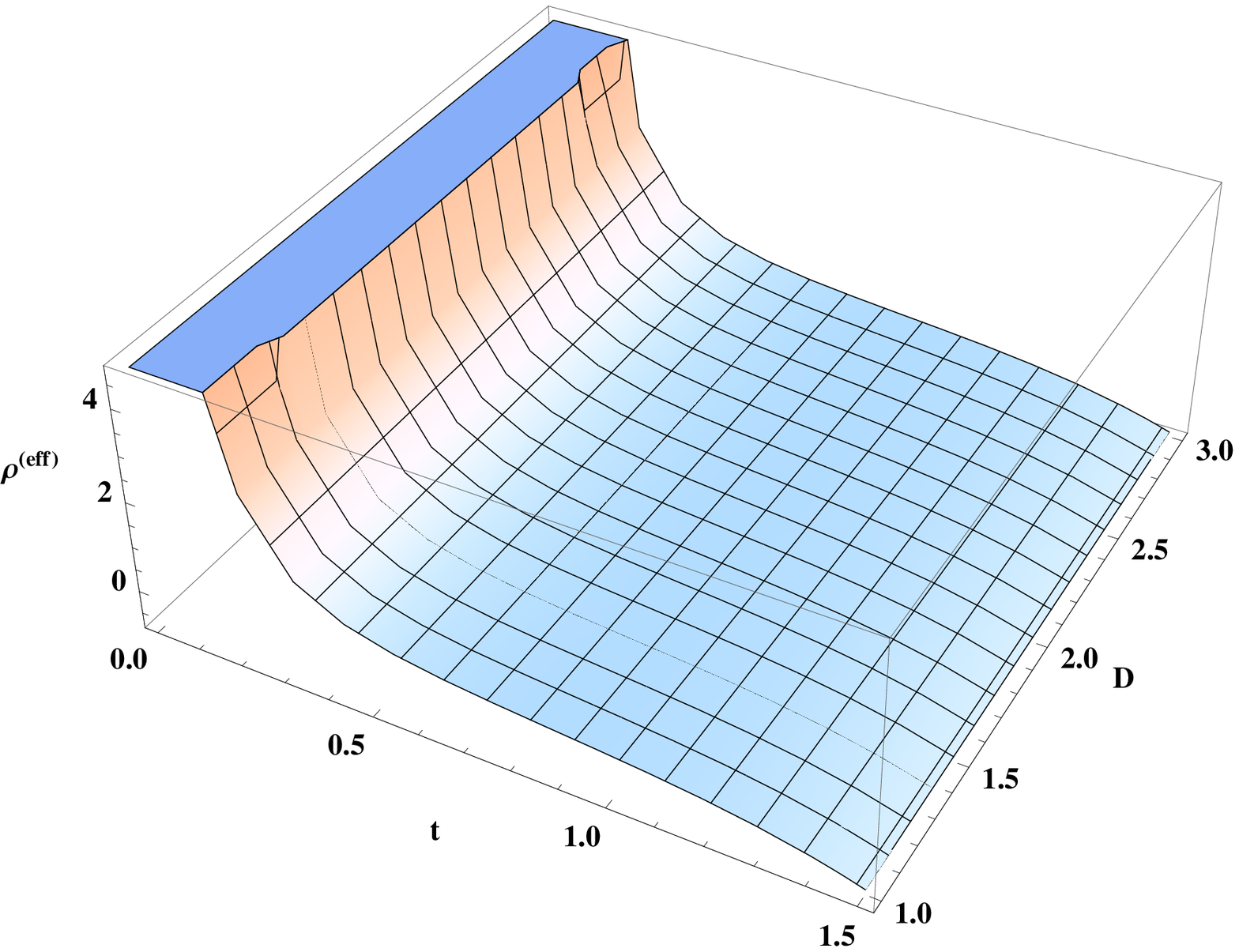}
\includegraphics[width=7.5cm]{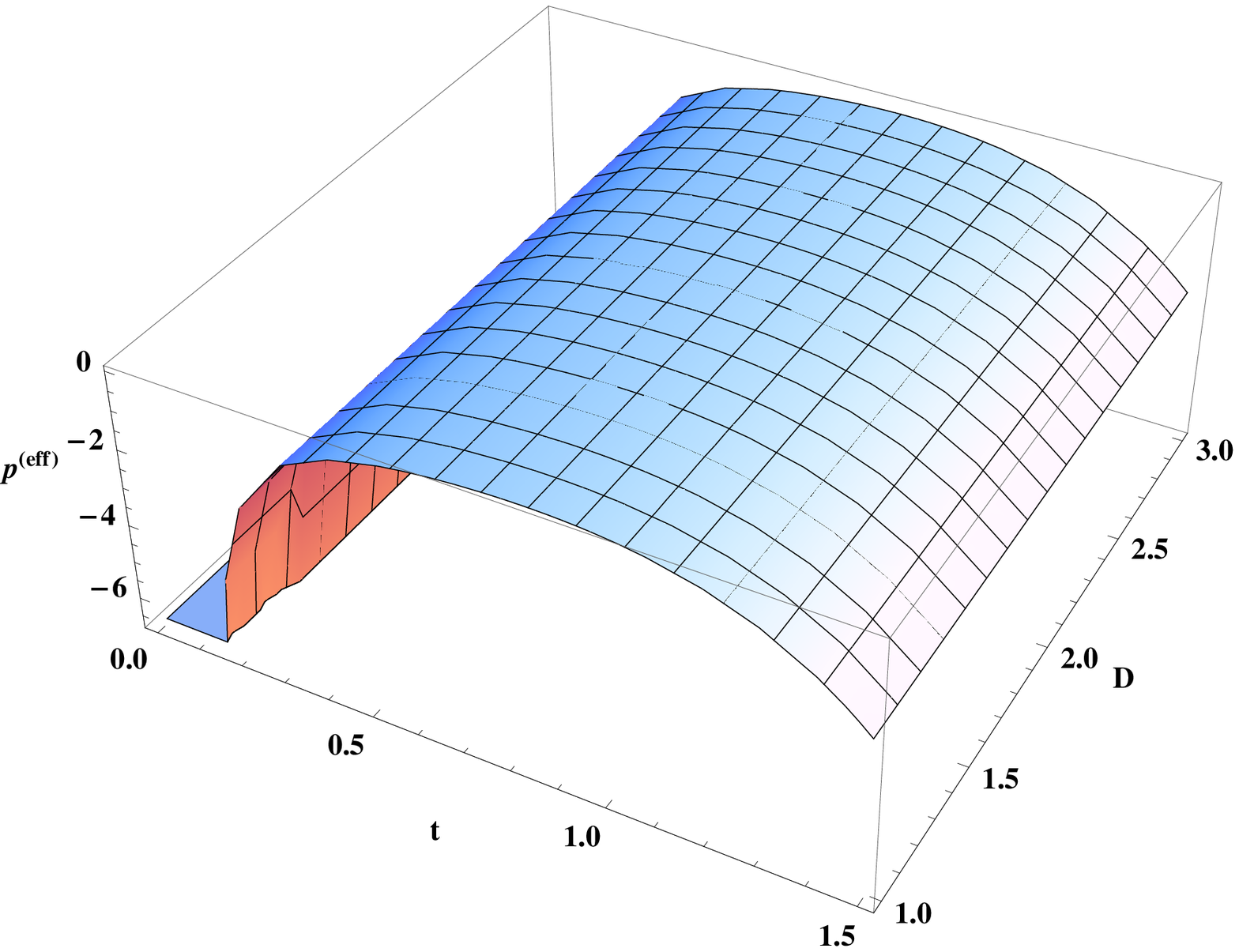}
\end{tabular}
\caption{Variation of $\rho^{(eff)}$ (left panel) and $~p^{(eff)}$ (right panel) versus time for $b_{1}=.3$, $b_{2}=.7$, $b_{3}=-1$, $\lambda=0.25$, $m=1.25$, $n=0.75$.}
\end{figure*}
\begin{figure*}[thbp]
\begin{tabular}{rl}
\includegraphics[width=7.5cm]{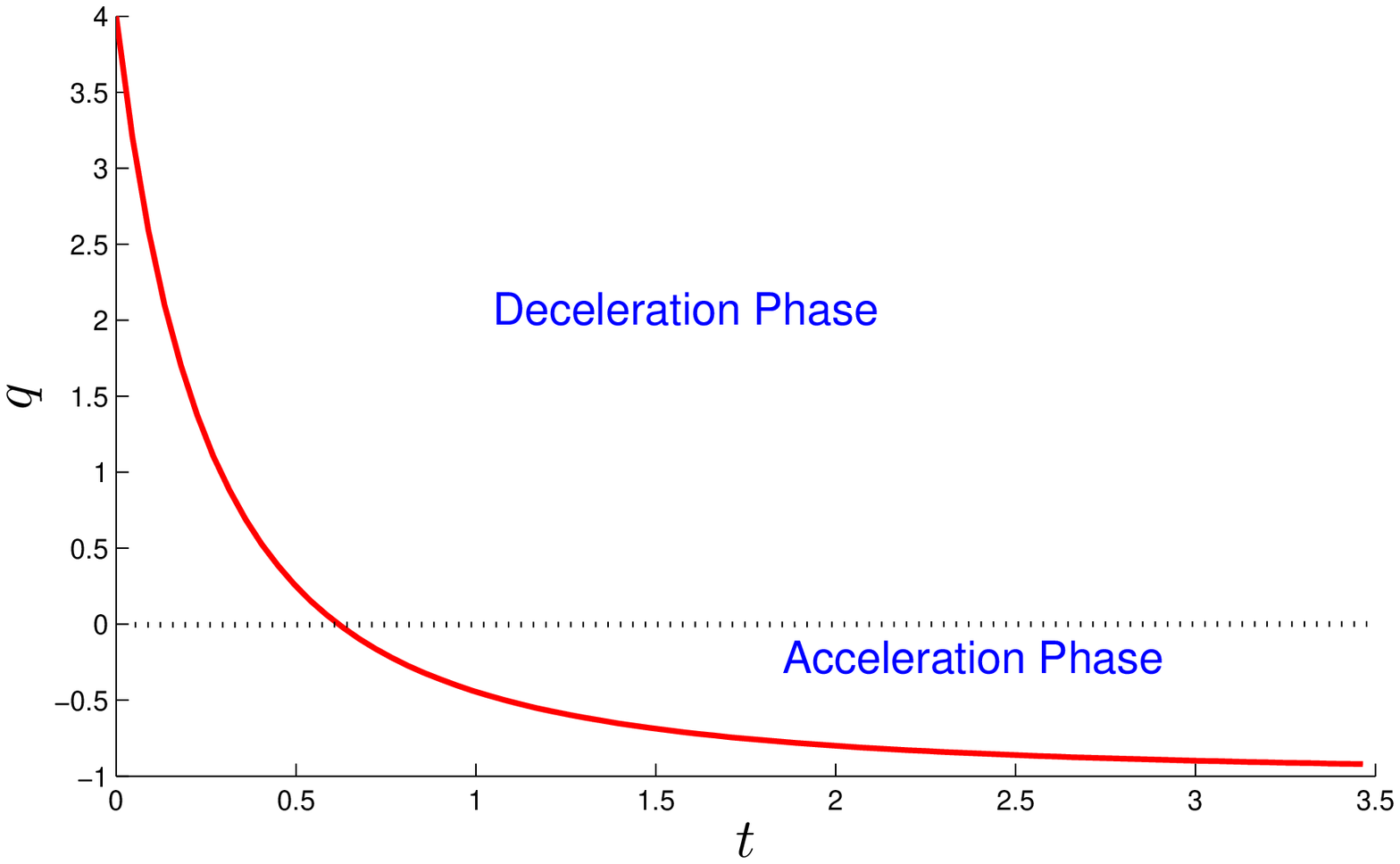}
\includegraphics[width=7.5cm]{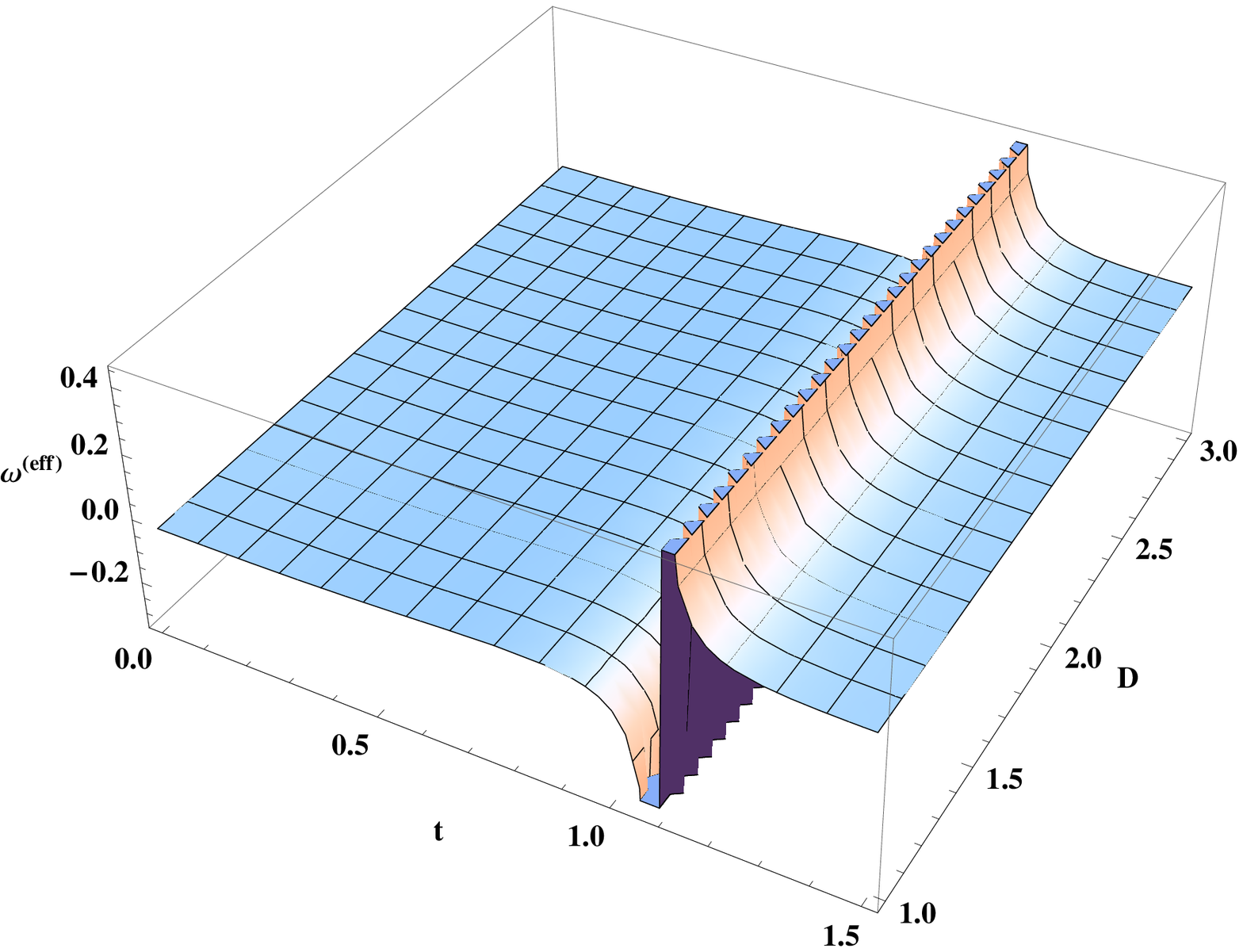}
\end{tabular}
\caption{The behavior of deceleration parameter (left panel) and $\omega^{\;eff}$ (right panel) versus time for $b_{1}=.3$, $b_{2}=.7$, $b_{3}=-1$, $\lambda=0.25$, $m=1.25$, $n=0.75$.}
\end{figure*}
\begin{figure*}[thbp]
\begin{tabular}{rl}
\includegraphics[width=8.5cm]{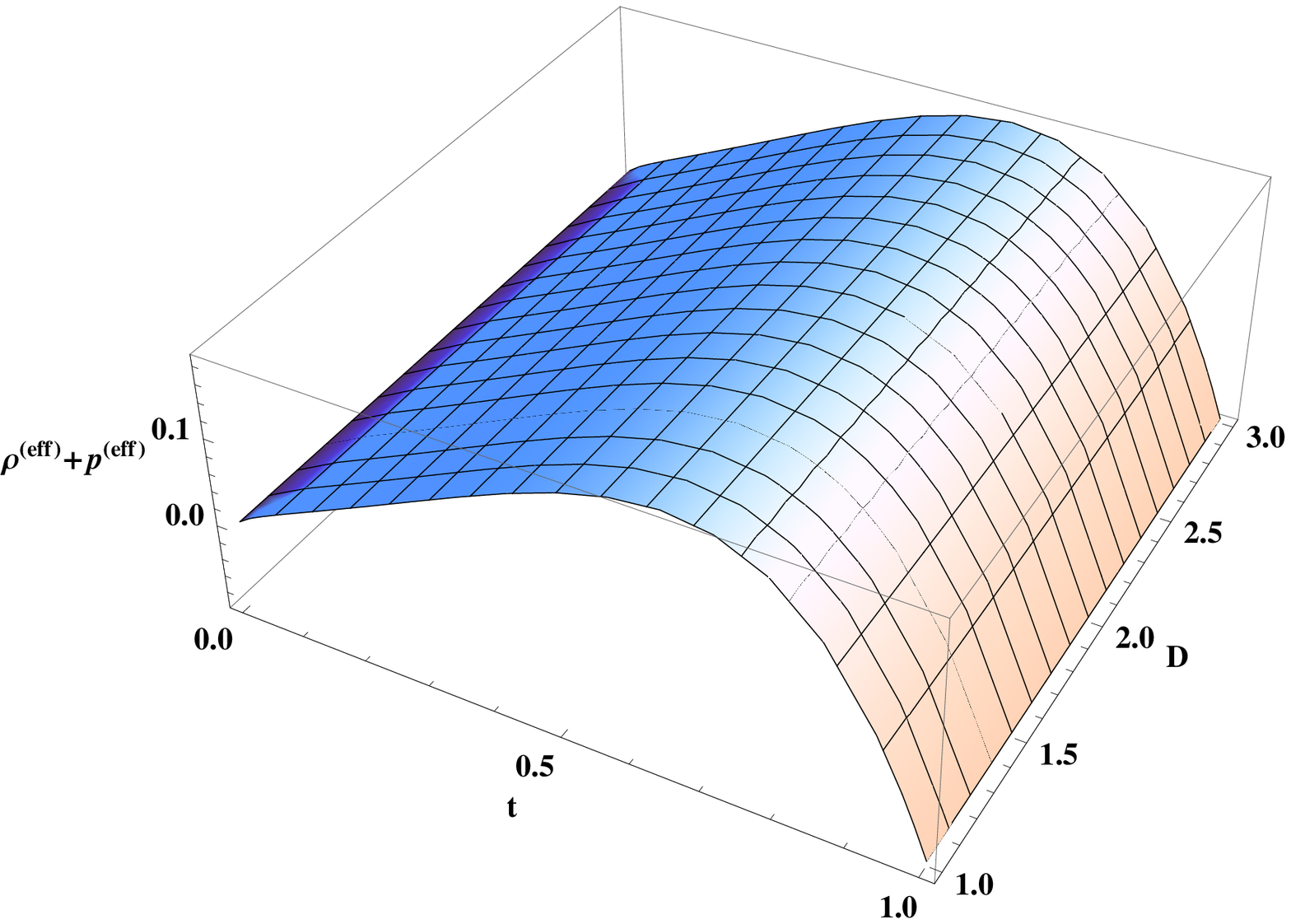}
\end{tabular}
\caption{Validation of WEC for $b_{1}=.3$, $b_{2}=.7$, $b_{3}=-1$, $\lambda=0.25$, $m=1.25$, $n=0.75$}
\end{figure*}
\begin{figure*}[thbp]
\begin{tabular}{rl}
\includegraphics[width=8.5cm]{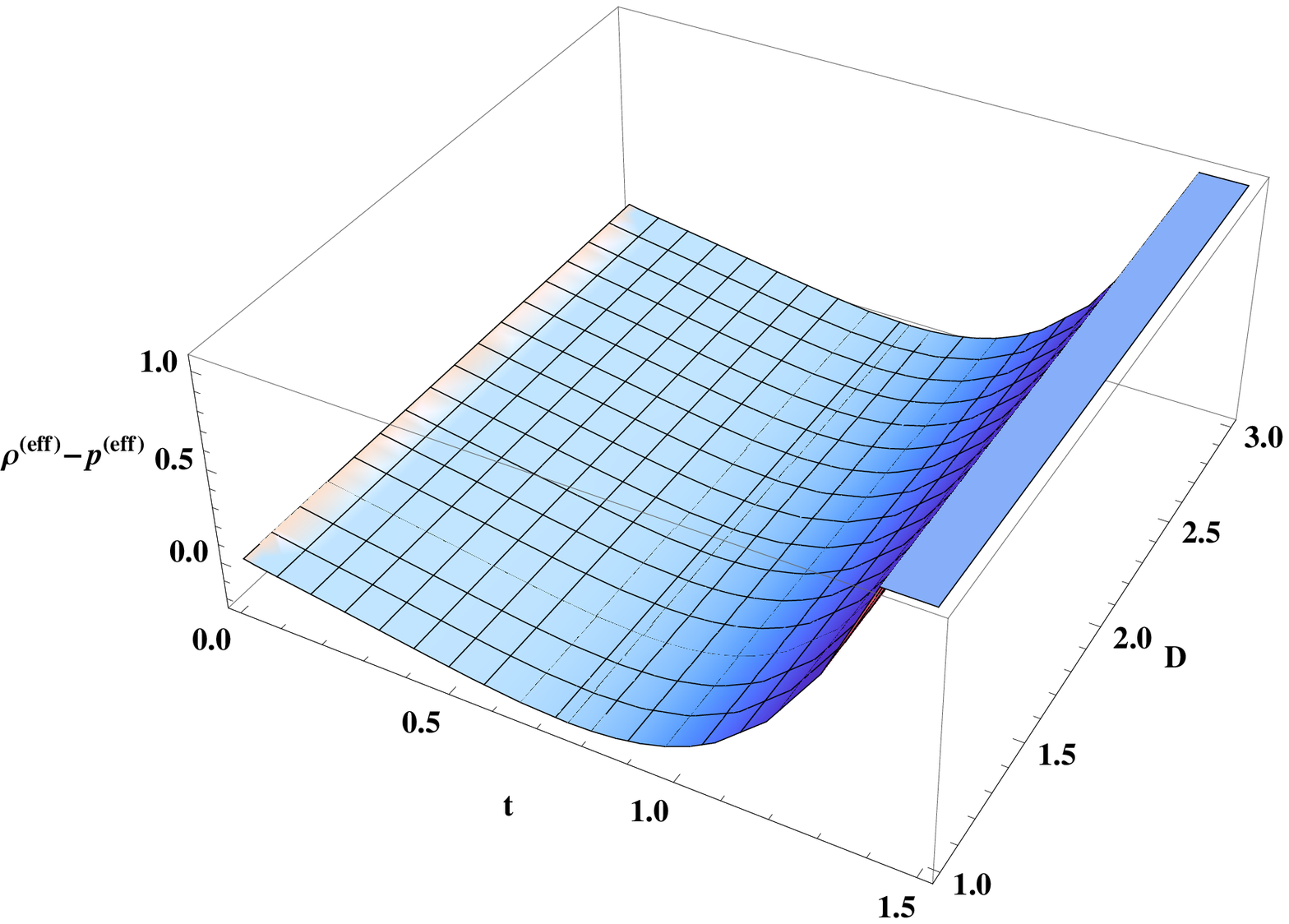}
\end{tabular}
\caption{Validation of DEC for $b_{1}=.3$, $b_{2}=.7$, $b_{3}=-1$, $\lambda=0.25$, $m=1.25$, $n=0.75$}
\end{figure*}
\begin{figure*}[thbp]
\begin{tabular}{rl}
\includegraphics[width=8.5cm]{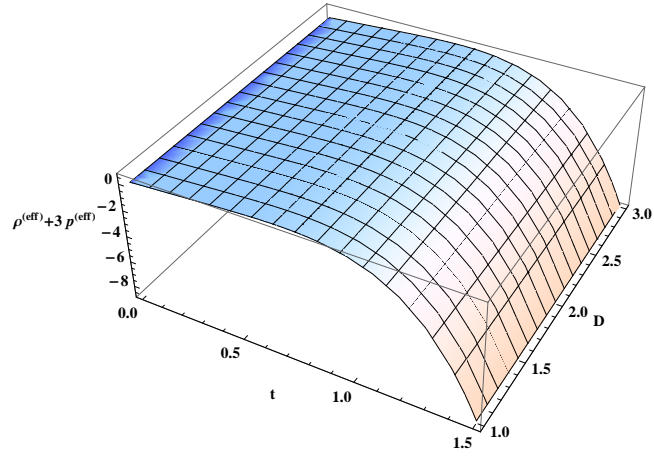}
\end{tabular}
\caption{Violation of SEC for $b_{1}=.3$, $b_{2}=.7$, $b_{3}=-1$, $\lambda=0.25$, $m=1.25$, $n=0.75$}
\end{figure*}
\begin{center}
\begin{figure*}[thbp]
\includegraphics[width=8.0cm]{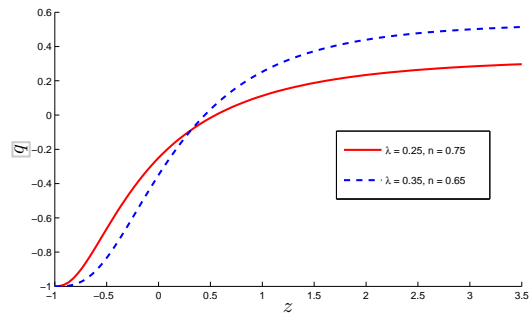}
\caption{Evolution of DP with $z$ for different values of $\lambda$ and $n$.}
\end{figure*}
\end{center}
\begin{figure*}[thbp]
\begin{tabular}{rl}
\includegraphics[width=7.5cm]{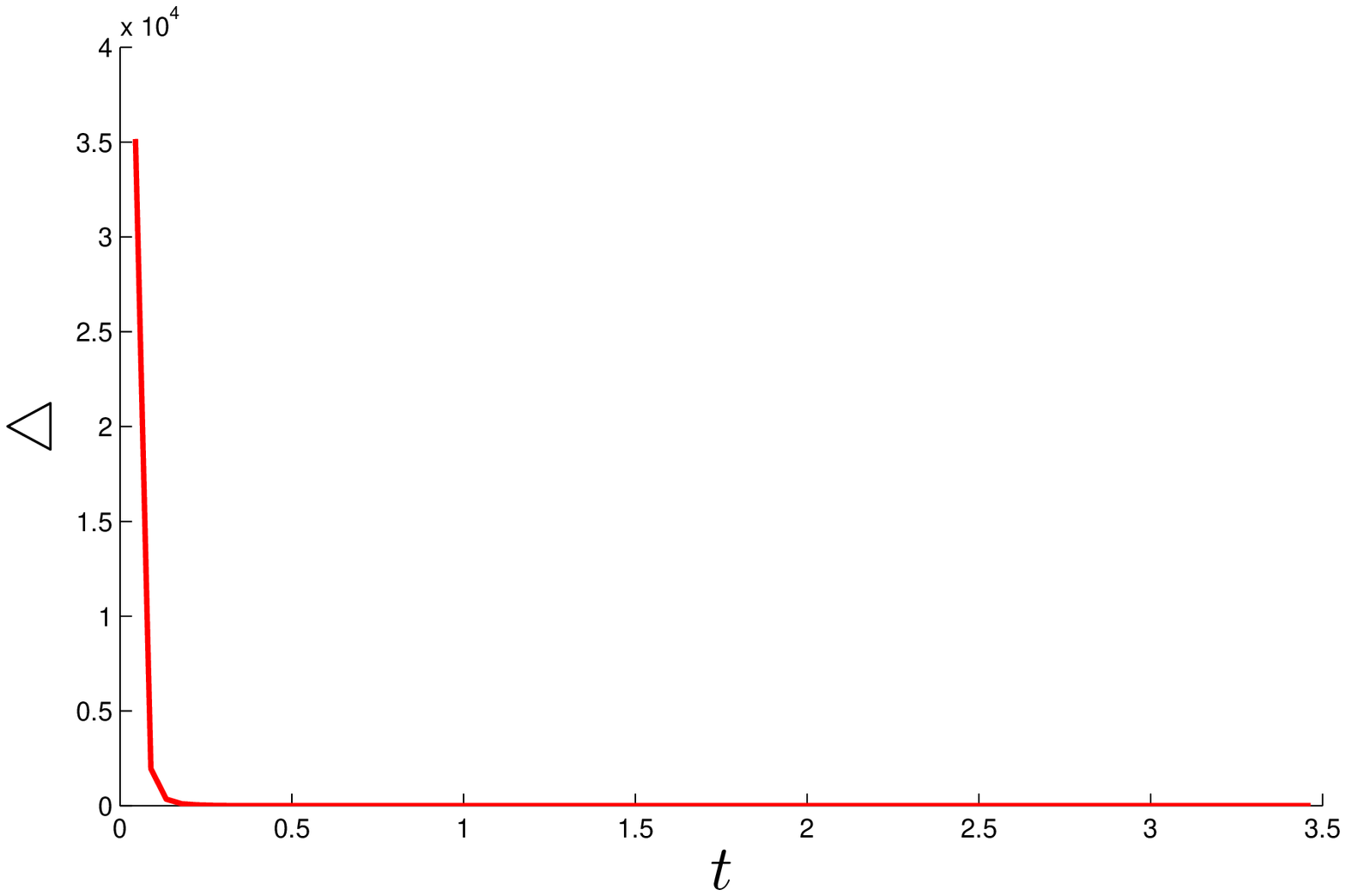}
\includegraphics[width=7.5cm]{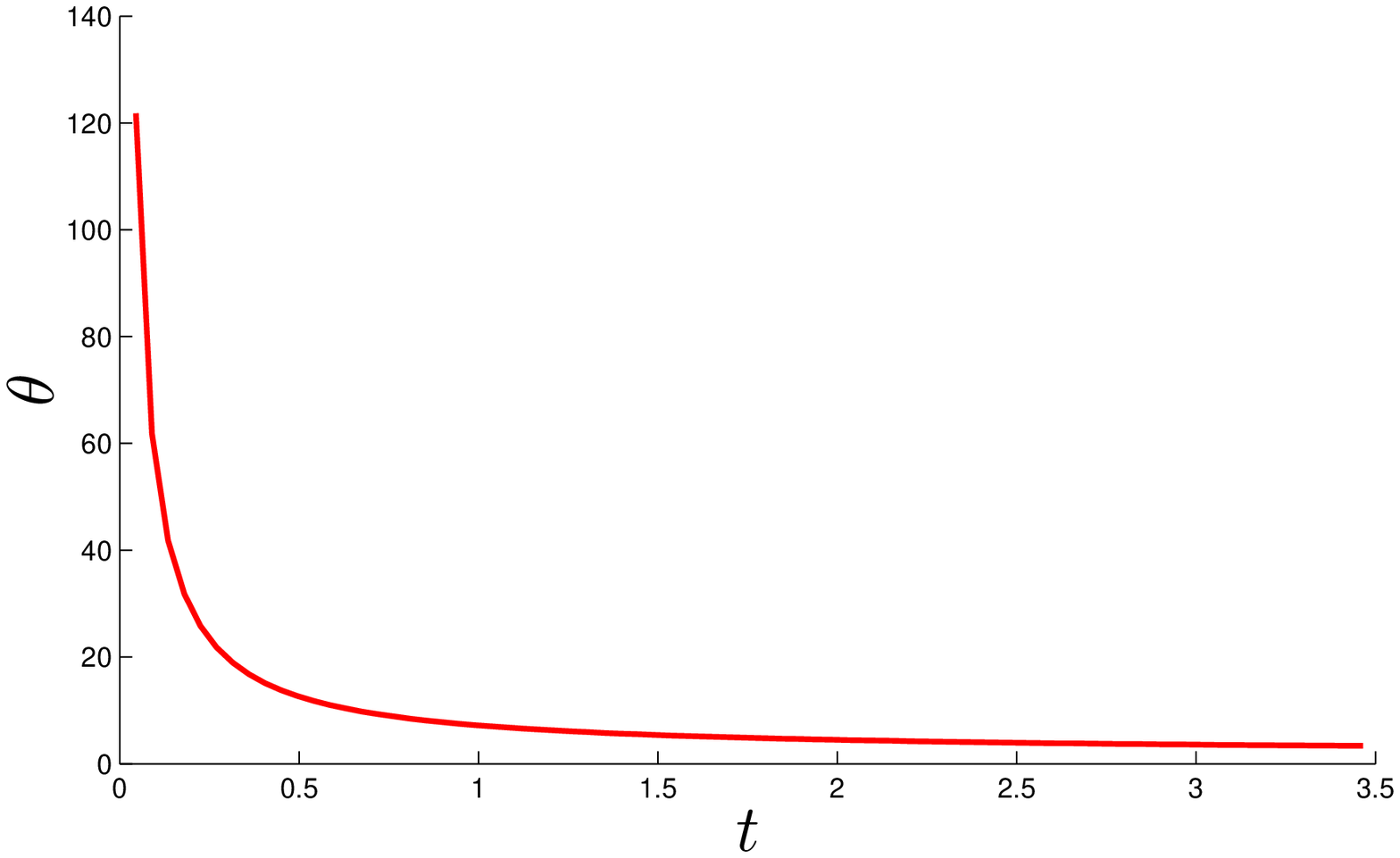}\\
\includegraphics[width=7.5cm]{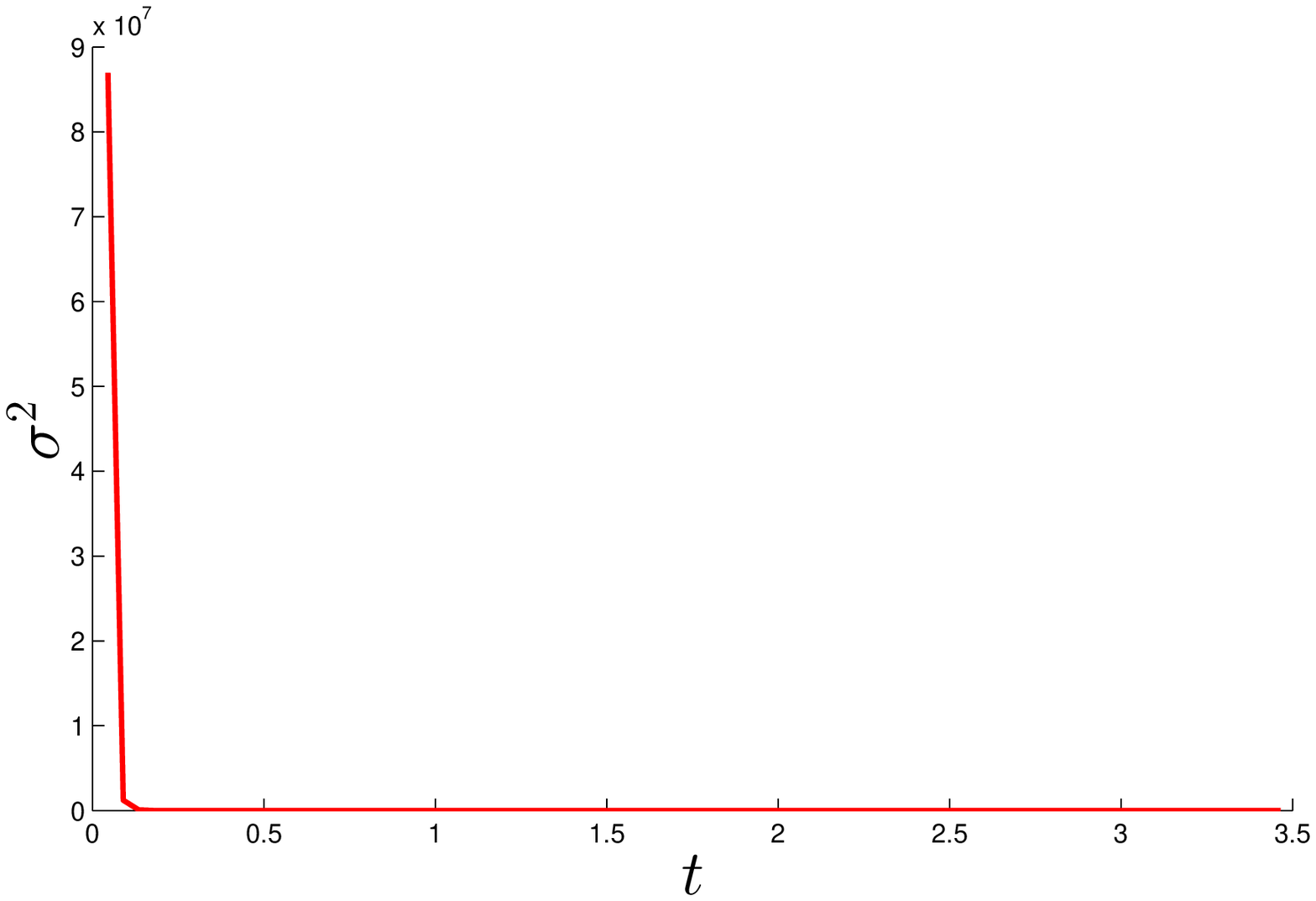}
\includegraphics[width=7.5cm]{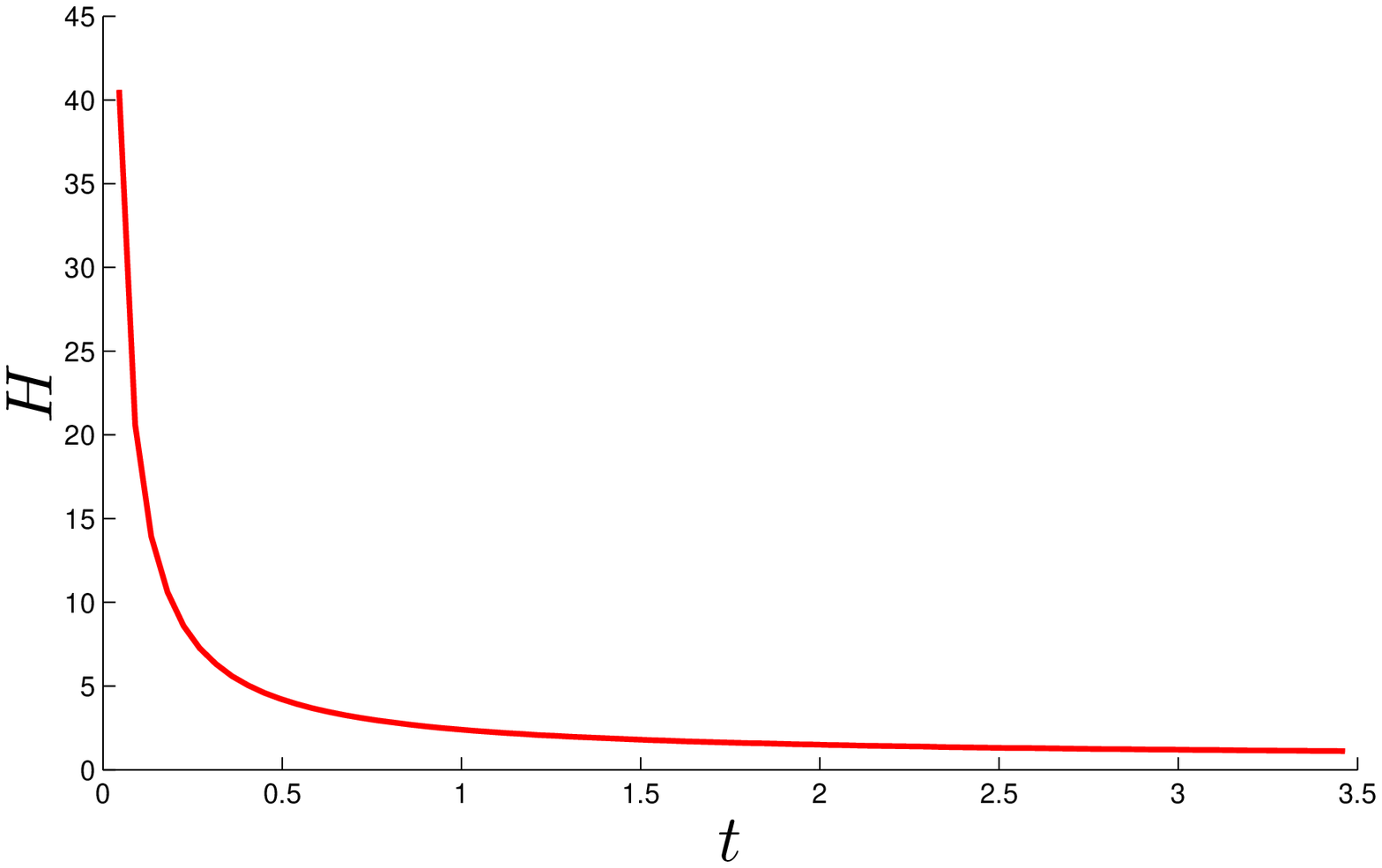}
\end{tabular}
\caption{Variation of $\triangle$ (left panel), $\theta$ (upper right panel), $\sigma^{2}$ (lower left panel) and $H$ (lower right panel) versus time for $b_{1}=.3$, $b_{2}=.7$, $b_{3}=-1$, $\lambda=0.25$, $m=1.25$, $n=0.75$.}
\end{figure*}
The line-element (\ref{spacetime}) and field equation (\ref{FRT2}) lead the following equations:
\begin{equation}
\label{fe1}
 \frac{\ddot{B}}{B}+\frac{\ddot{C}}{C}+\frac{\dot{B}\dot{C}}{BC} = -8\pi p^{(eff)}
\end{equation}
\begin{equation}
\label{fe2}
 \frac{\ddot{C}}{C}+\frac{\ddot{A}}{A}+\frac{\dot{A}\dot{C}}{AC} = -8\pi p^{(eff)}
\end{equation}
\begin{equation}
\label{fe3}
 \frac{\ddot{A}}{A}+\frac{\ddot{B}}{B}+\frac{\dot{A}\dot{B}}{AB} = -8\pi p^{(eff)}
\end{equation}
\begin{equation}
\label{fe4}
 \frac{\dot{A}\dot{B}}{AB}+\frac{\dot{B}\dot{C}}{BC}+\frac{\dot{C}\dot{A}}{CA} = 8\pi \rho^{(eff)}
\end{equation}
Here, $\rho^{(eff)} = \rho + \rho^{(DE)} = \rho -\frac{3\alpha}{8\pi}\left(\frac{\ddot{a}}{a}+
\frac{\dot{a}^{2}}{a^{2}}\right)(3\rho - 7p)$,
$p^{(eff)} = p + p^{(DE)} = p +\frac{9\alpha}{8\pi}\left(\frac{\ddot{a}}{a}+
\frac{\dot{a}^{2}}{a^{2}}\right)(\rho - 3p)$ and $a= (ABC)^{\frac{1}{3}}$ is average scale factor.\\

The above equations (\ref{fe1})$-$(\ref{fe4}) can also be written as
\begin{equation}
 \label{ge}
\frac{(ABC)^{\ddot{}}}{ABC} = 12\pi(\rho^{(eff)}-p^{(eff)})
\end{equation}

\subsection{Solution of field equations \& physical parameters}
The generalized HEL form of scale factor \cite{Ozgur/2014,Yadav/2013,Yadav/2016} is given by
\begin{equation}
\label{scale}
a=[t^{n}exp(\lambda t)]^{\frac{1}{m}}
\end{equation} 
where, $m$, $n$ and $\lambda$ are positive constants.\\

Solving equations (\ref{fe1})$-$(\ref{fe4}) with equation (\ref{scale}), we obtain
\begin{equation}
\label{A}
A(t)=a_{1}[t^{n}exp(\lambda t)]^{\frac{1}{m}}exp\left[b_{1}\int \left(t^{n}exp(\lambda t)\right)^{-\frac{3}{m}}dt\right]
\end{equation}
\begin{equation}
\label{B}
B(t)=a_{2}[t^{n}exp(\lambda t)]^{\frac{1}{m}}exp\left[b_{2}\int \left(t^{n}exp(\lambda t)\right)^{-\frac{3}{m}}dt\right]
\end{equation}
\begin{equation}
\label{C}
C(t)=a_{3}[t^{n}exp(\lambda t)]^{\frac{1}{m}}exp\left[b_{3}\int \left(t^{n}exp(\lambda t)\right)^{-\frac{3}{m}}dt\right]
\end{equation}
where $a_{1}$, $a_{2}$, $a_{3}$, $b_{1}$, $b_{2}$ \& $b_{3}$ are constants that fulfill the following requirements:\\
$a_{1}a_{2}a_{3} = 1$ \& 
$b_{1}+b_{2}+b_{3} = 0$.\\
The deceleration parameter of derived model is given by
\begin{equation}
\label{dp}
q = \frac{mn}{(n+\lambda t)^{2}}-1
\end{equation}
Solving equations (\ref{fe1})$-$(\ref{fe4}) and (\ref{A})$-$(\ref{C}), the expressions for $\rho^{(eff)}$ and 
$p^{(eff)}$ are respectively read as
\begin{equation}
\label{rho}
\begin{array}{ll}
8\pi\rho^{(eff)} = \frac{3}{m^{2}}\left(\lambda + \frac{m}{t}\right)^{2}+ (b_{1}b_{2}+b_{2}b_{3}+b_{3}b_{1}) \times\\
\\
\,\,\,\,\,\,\,\,\,\,\,\,\,\,\,\,\,\,\,\,\,\,\,\,\,\,\,\,\,\,\,\,\,\,\,\,\,\,\,\,
[t^{n}exp(\lambda t)]^{-\frac{6}{m}}.
\end{array}
\end{equation}
\begin{equation}
\label{p}
\begin{array}{ll}
8\pi p^{(eff)} = 
\frac{6b_{1}}{m}(t^{n}exp(\lambda t)^{-\frac{3}{m}}\left[3(\lambda+\frac{m}{t})-3t^{3}exp(\lambda t)\right]\\
\\
\,\,\,\,\,\,\,\,\,\,
-(b_{1}b_{2}+b_{2}b_{3}+b_{3}b_{1})[t^{n}exp(\lambda t)]^{-\frac{6}{m}}+\frac{2n}{mt^{2}}\\
\\
\,\,\,\,\,\,\,\,\,\,\,\,\,\,\,\,\,\,\,\,\,\,\,\,\,\,\,
-\frac{3}{m^{2}}\left(\lambda + \frac{m}{t}\right)^{2}.
\end{array}
\end{equation}
\section{Physical consequences of the model \& energy conditions}
We observe that the equations (\ref{A})$-$(\ref{p}), identically satisfy equation (\ref{ge}). Hence the solutions obtained in this paper are exact as well as important for describing the dynamics of physical universe. The Hubble's parameter is computed as $H = \frac{3}{m}\left(\frac{n}{t}+\lambda\right)$. In the derived model, it has been seen also seen that the expansion scalar $(\theta)$ is proportional to Hubble's parameter $(H)$ and $V = a^{3}$.\\
The anisotropy parameter and shear scalar are read as
\begin{equation}
\label{an}
\triangle = \frac{m^{2}(b_{1}^{2}+b_{2}^{2}+b_{3}^{2})}{27\left(\frac{n}{t}+\lambda\right)}[t^{n}exp(\lambda t)]^{-\frac{6}{n}}
\end{equation}
\begin{equation}
\label{shear}
\sigma^{2} = \frac{(b_{1}^{2}+b_{2}^{2}+b_{3}^{2})}{2}[t^{n}exp(\lambda t)]^{-\frac{6}{n}}
\end{equation} 

The anisotropy parameter and shear scalar decrease with time that matches with the properties of realistic universe.\\

The behaviour of $\rho^{(eff)}$ \& $p^{(eff)}$ and deceleration parameter $(q)$ have been graphed in $Figs.~1~-~2$. 
It is evident that $\rho^{(eff)}$ decreases with time and finally drops to zero for long time and $p^{(eff)}$ is negative throughout the evolution of universe. This behaviour of $\rho^{(eff)}$ and $p^{(eff)}$ match with observed universe. Thus the derived model may able to explain the dynamics of accelerating universe without possible contribution of dark energy/dark matter in $f(R,T)$ gravity. From the right panel of $Fig.~1$, one may note that at  beginning the DP was positive and universe evolves with deceleration but after some time $q$ becomes negative which have consistency with the model of accelerating universe. At late time the value of $q$ approaches to $-1$ which shows the fastest rate of expansion of universe at late time.\\       

The $\omega^{(eff)} = p^{(eff)}/\rho^{(eff)}$ and single plot of energy conditions have been graphed in $Figs.~ 2 - ~ 5$. 
From the left panel of $Fig.~2$, it has been noticed that the weak energy condition (WEC) and dominant energy condition (DEC) have been satisfied in the derived model but strong energy condition (SEC) is  violated which ensures that anti-gravitational effect - may be one of the possible cause of acceleration. The right panel of $Fig.~2$ depicts the behaviour of equation of state parameter $(\omega^{(eff)})$ versus time. For accelerating universe the value of $\omega^{(eff)}$ must be lies in between -1 and 0. In the simplest case, $\omega^{(eff)} = -1$ generates the $\Lambda$CDM model while phantom model and quintessence model also arise when $\omega^{(eff)} < -1$ and $\omega^{(eff)} > - 1$ respectively. For our model, we obtain present accepted numerical value of $\omega^{(eff)}$ as $-1$ for longer times.\\ 
From equation (\ref{scale}), one may express the scale factor in the term of redshift $(z)$ by taking into account the present value of scale factor i.e. $a_{0} = 1$, as following
\begin{equation}
\label{redshift}
t=\frac{n}{\lambda}W\left[\frac{\lambda}{n}\left(\frac{1}{1+z}\right)\right]^{\frac{m}{n}}
\end{equation} 
where $W$ denotes the $W$-function or product logarithm.\\

$Fig.~6$ depicts the behaviour of DP with redshift for different value of $\lambda$ and $n$. Here we choose the values of constant $\lambda$ and $n$ in agreement with the observational constraints reported in \cite{Ozgur/2014}. It is worth to note that the present values of q (i.e. at z = 0) are $-0.25$ and $-0.35$ for $\lambda = 0.25$, $n = 0.75$ and $\lambda = 0.35$, $n = 0.65$ respectively. These values are in agreement with recent observational data \cite{Hinshaw/2013}. From $Fig.~3$, one can check the value of redshift at which the universe transit itself decelerated phase to accelerated expansion. The transition of universe occurs at $z_{tr} = 0.55,~ 0.44$ corresponding to the $(\lambda, n) \sim $ $(0.25, 0.75)$ and $(0.35, 0.75)$ respectively. The WMAP observations \cite{Hinshaw/2013} favors the value of $z_{tr}$ obtained for $(\lambda, n) \sim $ $(0.25, 0.75)$ that is why we have graphed the other physical parameters by taking $\lambda = 0.25$ and $n = 0.75$.\\

\section {Conclusion}
In the present work, we have searched the existence of transitioning scenario of Bianchi-I universe with $f\left(R,T\right) = f_{1}(R) + f_{2}(R)f_{3}(T)$. In particular the exact solution of Einstein's field equation have been obtained by applying generalized hybrid expansion law of scale factor and functional forms $f_{1}(R) = f_{2}(R) = R$ and $f_{3}(T) = \alpha T$ with $\alpha$ being constant. From left panel of $Fig.~1$, one can observe that the effective energy density is positive and decreasing function of time while the effective pressure is negative through out the evolving process of universe. The validation of energy conditions have been graphed in $Figs.~3~-~5$. In the derived model the validation of energy conditions excepts SEC shows that the accelerating universe must violate SEC \cite{Barcelo/2002}. The behaviour of anisotropy parameter $(\triangle)$, expansion scalar $(\theta)$, shear scalar $(\sigma^{2})$ and Hubble's parameters $(H)$ have been graphed in $Fig.~7$. The parameters $H$, $\theta$, $\triangle$ and $\sigma^{2}$ start with extremely large values and continue to decrease with passage of time which mimic the present scenario of universe. \\
 
It is worth to note that for $m = 1$ and $A(t) = B(t) = C(t) = a(t)$, our model reproduces the result obtained in Ref. \cite{Moraes/2017}. Thus our model generalizes the solution obtained by Moraes and Sahoo \cite{Moraes/2017} and evoke the theory of $f(R,T)$ gravity in anisotropic space-time. Further we analyze that for $\lambda = 0$, the dynamics of derived model is governed by power law which gives singular universe with big bang singularity at $t = 0$. Similarly for $n = 0$, the derived model gives the dynamics of singularity free universe. The DP (q) is found to be positive in early universe and it becomes negative at late time. The present value of $q~(z =0)$ is estimated as $-0.25$. This value of DP matches with observational value of q at present epoch, reported in Ref. \cite{Hinshaw/2013}. The right panel of $Fig.~2$ shows the dynamics of $\omega^{(eff)}$ with passage of time. The evolving range of $\omega^{(eff)}$ in our model is agreed with previous results \cite{Yadav/2011,Yadav/2011a,Kumar/2011,Saha/2012a}. \\      


\begin{thebibliography}{0}

\bibitem{Harko/2011} T. Harko, F.S.N. Lobo, S. Nojiri and S.D. Odintsov, $f(R,T)$ gravity \textit{Phys. Rev. D} \textbf{84} (2011) 024020.

\bibitem{Riess/1998} A. G. Riess et al., Observational Evidence from Supernovae for an Accelerating Universe and a Cosmological Constant \textit{Astron. J.} \textbf{116} (1998) 1009.

\bibitem{Perlmutter/1999} S. Perlmutter et al., Measurements of $\Omega$ and $\Lambda$ from 42 High-Redshift Supernovae Astrophys. J. \textbf{517} (1999) 565.

\bibitem{Kumar/2017} S. Kumar, R. C. Nunes, Observational constraints on dark matter - dark energy scattering cross section \textit{Eur. Phys. J. C.} \textbf{77} (2017) 734.

\bibitem{Dil/2017} E. Dil, Cosmology of q-deformed dark matter and dark energy \textit{Phys. Dark Univ.} \textbf{16} (2017) 1.

\bibitem{Behrouz/2017} N. Behrouz et al, Interacting quintom dark energy with Nonminimal Derivative Coupling \textit{Phys. Dark Univ.} \textbf{15} (2017) 72.

\bibitem{Germani/2017} C. Germani, Initial conditions for the Galileon dark energy, \textit{Phys. Dark Univ.} \textbf{15} (2017) 1. 

\bibitem{Jennen/2016} H. Jennen, J. G. Pereira, Dark energy as a kinematic effect, \textit{Phys. Dark Univ.} \textbf{11} (2016) 49.

\bibitem{Moraes/2017} P.H.R.S. Moraes, P.K. Sahoo, The simplest non-minimal matter-geometry coupling in the $f(R,T)$ cosmology \textit{Eur. Phys. J. C} \textbf{77} (2017) 480.

\bibitem{Yadav/2014} A.K. Yadav, Bianchi-V string cosmology with power law expansion in $f(R,T)$ gravity \textit{Euro Phys. J. Plus} \textbf{129} (2014) 194.

\bibitem{Yadav/2018} A. K. Yadav, A. T. Ali, Invariant Bianchi type I models in $f(R,T)$ gravity \textit{Int. J. Geom. Methods in Mod. Phys.} doi.org/10.1142/S0219887818500263 (2017)

\bibitem{Yadav/2015} A. K. Yadav, P. K. Srivastava, L. Yadav, Hybrid Expansion Law for Dark Energy Dominated
Universe in f (R,T) Gravity \textit{Int. J. Theor. Phys.} \textbf{54} (2015) 1671.

\bibitem{Yadav/2013} A. K. Yadav, A. Sharma, A transitioning universe with time varying G and decaying $\Lambda$ \textit{Research in Astron. Astrophys.} \textbf{13}, (2013) 501.

\bibitem{Singh/2015} V. Singh, C. P. Singh, Friedmann Cosmology with Matter Creation in Modified $f(R,T)$ Gravity \textit{Int. J. Theor. Phys.} \textbf{55} (2015) 1257.

\bibitem{Parker/2014} L. Parker, Quantized Fields and Particle Creation in Expanding Universes \textit{Phys. Rev. D} \textbf{3} (2071) 2546-2546.

\bibitem{Myrazakulov/2012} R. Myrzakulov, FRW Cosmology in $f(R,T)$ gravity \textit{Eur. Phys. J. C} \textbf{72} (2012) 2203. 

\bibitem{Houndjo/2012} M. J. S. Houndjo, Reconstruction of $f(R,T)$ gravity describing matter dominated and accelerated phases \textit{Int. J. Mod. Phys. D} \textbf{21} (2012) 1250003.

\bibitem{Jamil/2012} M. Jamil, D. Momeni, M. Raza, R. Mryzakulov, Reconstruction of some cosmological models in $f(R,T)$ cosmology \textit{Eur. Phys. J. C} \textbf{72} (2012) 1999. 

\bibitem{Kiani/2014} F. Kiani, K. Nozari, Energy conditions in $f(T,\theta)$ gravity and compatibility with a stable de Sitter solution \textit{Phys. Lett. B} \textbf{728} (2014) 554-561.

\bibitem{Zubair/2016} M. Zubair, S. Waheed, Y. Ahmad, Static Spherically Symmetric Wormholes in $f(R,T)$ Gravity \textit{Eur. Phys. J. C.} \textbf{76} (2016) 444. 

\bibitem{Yousaf/2017} Z. Yousaf, M Ilyas, M. Z. Bhatti, Influence of modification of gravity on spherical wormhole models \textit{Mod. Phys. Lett. A} \textbf{32} (2017) 1750163.
 
\bibitem{Moraes/2017r} P. H. R. S. Moraes, R. A. C. Correa, R. V. Lobato, Analytical general solutions for static wormholes in $f(R,T)$ gravity \textit{JCAP} \textbf{07} (2017) 029.

\bibitem{MoraesSahoo/2017} P. H. R. S. Moraes, P.K. Sahoo, Modelling wormholes in $f(R,T)$ gravity \textit{Phys. Rev. D} \textbf{96} (2017) 044038.

\bibitem{Das/2016} A. Das, F. Rahaman, B. K. Guha, S. Ray, Compact stars in $f(R,T)$ gravity \textit{Eur. Phys. J. C} \textbf{76} (2016) 654  

\bibitem{Sharif/2014} M. Sharif, Z. Yousaf, Dynamical analysis of self-gravitating stars in $f(R, T)$ gravity \textit{Astrophys. Space Sc.} \textbf{354} (2014) 471.  

\bibitem{Shabani/2014} H. Shabani, M. Farhoudi, Cosmological and Solar System Consequences of $f(R, T)$ Gravity Models \textit{Phys. Rev. D} \textbf{90} (2014) 044031.

\bibitem{Shen/2014} M. Shen, L. Zhao, Oscillating Quintom Model with Time Periodic Varying Deceleration Parameter \textit{Chin. Phys. Lett.} \textbf{31} (2014) 010401

\bibitem{Akarsu/2010} O. Akarsu, C. B. Kilinc, LRS Bianchi type I models with anisotropic dark energy and constant deceleration parameter \textit{Gen. Relativ. Grav.} \textbf{42} (2010) 119-140.

\bibitem{Kumar/2011} S. Kumar, C. P. Singh, Anisotropic dark energy models with constant deceleration parameter \textit{Gen. Relativ. Grav.} \textbf{43} (2011) 1427-1442.

\bibitem{Yadav/2012} A. K. Yadav, B. Saha, LRS Bianchi-I anisotropic cosmological model with dominance of dark energy \textit{Astrophys. Space Sc.} \textbf{337} (2012) 759–765.

\bibitem{Saha/2004} B. Saha, T. Boyadjiev, Bianchi type-I cosmology with scalar and spinor fields \textit{Phys. Rev. D} \textbf{69} (2004) 124010.

\bibitem{Yadav/2016} A. K. Yadav, A transitioning universe with anisotropic dark energy \textit{Astrophys. Space Sc.} \textbf{361} (2016) 276.

\bibitem{Barcelo/2002} C. Barcelo, M. Visser, Twilight for the energy conditions \textit{Int. J. Mod. Phys. D} \textbf{11} (2002) 1553.

\bibitem{Ozgur/2014} A. Ozgur et al, Cosmology with hybrid expansion law: scalar field reconstruction of
cosmic history and observational constraints \textit{JCAP} \textbf{01} (2014) 022

\bibitem{Hinshaw/2013} G. Hinshaw et al, Nine-Year WMAP Observations:
Cosmological parameter results \textit{Astrophys. J.} \textbf{208} (2013) 19

\bibitem{Aygun/2016} S. Aygun, C. Aktas, I. Yilmaz, \textit{Astrophysics \& Space Sc.} \textbf{361} (2016) 380.

\bibitem{Moraes/2017a} P.H.R.S. Moraes, P.K. Sahoo, G. Ribeiro, R. A. C. Correa, A cosmological scenario from Starobinsky model with $f(R,T)$ formalism arXiv: 1712.07569 [gr-qc] (2017).

\bibitem{Zaregonbadi/2016} Zaregonbadi R et al, Dark matter from $f(R,T)$ gravity \textit{Phys. Rev. D} \textbf{94} (2016) 084052

\bibitem{Yadav/2011} A. K. Yadav, L. Yadav, Bianchi Type III Anisotropic Dark Energy Models with Constant Deceleration Parameter \textit{Int. J. Theor. Phys.} \textbf{50} (2011) 218.

\bibitem{Yadav/2011a} A. K. Yadav, Some Anisotropic Dark Energy Models in Bianchi Type-V Space-time \textit{Astrophys Space Sc.} \textbf{335} (2011) 565.

\bibitem{Saha/2012a} B. Saha, A. K. Yadav, Dark energy model with variable q and $\omega$ in LRS Bianchi-II space-time \textit{Astrophys Space Sc.} \textbf{341} (2012) 651.

\end{thebibliography}
\end{document}